\documentclass[preprint]{aastex}
\shortauthors{Chung et al.}
\shorttitle{A PLANETARY FEATURE IN CAUSTIC-CROSSING EVENTS}

\newcommand{\thetae}{\theta_{\rm E}}
\newcommand{\thetaeone}{\theta_{\rm {E,1}}}

\newcommand{\sbb}{s_{\rm b}}
\newcommand{\spp}{s_{\rm p}}
\newcommand{\qb}{q_{\rm b}}
\newcommand{\qp}{q_{\rm p}}
\newcommand{\Rb}{R_{\rm b}}
\newcommand{\Rp}{R_{\rm p}}
\newcommand{\sbhat}{{\hat s_{\rm b}}}
\newcommand{\delxib}{\Delta \xi_{\rm b}}
\newcommand{\deletab}{\Delta \eta_{\rm b}}
\newcommand{\delxip}{\Delta \xi_{\rm p}}
\newcommand{\deletap}{\Delta \eta_{\rm p}}

\begin{document}
\title{A planetary lensing feature in caustic-crossing high-magnification microlensing events}

\author{
Sun-Ju Chung\altaffilmark{1},
Kyu-Ha Hwang\altaffilmark{1},
Yoon-Hyun Ryu\altaffilmark{1,2}, and
Chung-Uk Lee\altaffilmark{1}
 }
\altaffiltext{1}
{Korea Astronomy and Space Science Institute, Hwaam-Dong,
Yuseong-Gu, Daejeon 305-348, Republic of Korea; sjchung@kasi.re.kr, kyuha@kasi.re.kr, yhryu@kasi.re.kr, leecu@kasi.re.kr}
\altaffiltext{2}
{Department of Astronomy and Atmospheric Sciences, Kyungpook National University, Daegu 702-701, Korea}


\begin{abstract}
Current microlensing follow-up observations focus on high-magnification events because of the high efficiency of planet detection.
However, central perturbations of high-magnification events caused by a planet can also be produced by a very close or a very wide binary companion, and the two kinds of central perturbations are not generally distinguished without time consuming detailed modeling (a planet-binary degeneracy).
Hence, it is important to resolve the planet-binary degeneracy that occurs in high-magnification events.
In this paper, we investigate caustic-crossing high-magnification events caused by a planet and a wide binary companion.
From this investigation, we find that because of the different magnification excess patterns inside the central caustics induced by the planet and the binary companion, the light curves of the caustic-crossing planetary-lensing events exhibit a feature that is discriminated from those of the caustic-crossing binary-lensing events, and the feature can be used to immediately distinguish between the planetary and binary companions.
The planetary-lensing feature appears in the interpeak region between the two peaks of the caustic-crossings.
The structure of the interpeak region for the planetary-lensing events is smooth and convex or boxy, whereas the structure for the binary-lensing events is smooth and concave.
We also investigate the effect of a finite background source star on the planetary-lensing feature in the caustic-crossing high-magnification events.
From this, we find that the convex-shaped interpeak structure appears in a certain range that changes with the mass ratio of the planet to the planet-hosting star.

\end{abstract}


\section{INTRODUCTION}

Survey and follow-up observations for the detection of a planet by using microlensing are being carried out toward the Galactic bulge.
Survey observations (OGLE: Udalski 2003; MOA: Bond et al. 2002) monitor a large area of sky and alert ongoing events by analyzing data in real time, while follow-up observations ($\mu$FUN: Dong et al. 2006; PLANET: Albrow et al. 2001; RoboNet: Burgdorf et al. 2007) intensively monitor the alerted events.
Because of the limited number of telescopes, follow-up observations cannot monitor all the alerted events.
Thus, current follow-up observations are focusing on high-magnification events.
High-magnification events for which the background source star passes close to the primary lens star are very sensitive for planet detection because the central caustic induced by a planet is formed near the primary star and thus produces central perturbations near the peak of the lensing light curves \citep{griest98}.
Hence, 8 of 13 extrasolar planets found by microlensing have been detected in high-magnification events (Udalski et al. (2005); Gould et al (2006); Gaudi et al. (2008); Bennett et al. (2008); Dong et al. (2009); Janczak et al. (2010); Miyake et al. (2011)).

However, central perturbations of high-magnification events can also be produced by a very close or a very wide binary companion \citep{gaudi97}.
This is because the size of the central caustic induced by the very close or the very wide binary companion is similar to that of the central caustic induced by a planet.
Because of this fact, high-magnification events with the central perturbations have a severe planet-binary degeneracy.
Fortunately, the planet-binary degeneracy can be resolved by time consuming detailed modeling because the light curves of events caused by the planet and the very close or the very wide binary companion are not perfectly identical (Albrow et al. 2002; Han \& Gaudi 2008).
In addition, the central perturbations caused by single planet systems can be produced by multiple planet systems and wide binary systems with a planet (Gaudi et al. 1998; Lee et al. 2008; Chung \& Park 2010). 
It occurs well in the limited case that the multiple planets and the planetary and binary companions in the individual systems are located at the same direction from the primary star.
Thus, there also exist degeneracies between the single planetary events and the multiple planetary events or the wide binary events with a planet.
As with the planet-binary degeneracy, the degeneracies are not perfect and thus can be resolved by the detailed modeling, but it is not easy to resolve them (Gaudi et al. (2008); Bennett et al. (2010)).
Considering the fact that current microlensing follow-up observations are focusing on high-magnification events to detect a planet, a simple diagnostic that can rapidly resolve the planet-binary degeneracy or the other degeneracies without the detailed modeling would be very helpful for a rapid understanding and analysis of lensing events being observed.

\citet{han08} studied double-peaked high-magnification events caused by planetary and wide binary companions.
They found a characteristic planetary feature in the double-peaked high-magnification events and showed that the characteristic feature can be used to immediately resolve the planet-binary degeneracy in observed high-magnification events. 
The characteristic planetary feature arises due to the different shape of the central caustics induced by the planetary and binary companions. 
The difference of the shapes of the two caustics affects both the perturbation regions inside and outside the caustic.
Han $\&$ Gaudi (2008) studied high-magnification events where central perturbations occur in the outside region of the two caustics.
We here study caustic-crossing high-magnification events where the central perturbations occur in the inside region of the two caustics.  

The paper is organized as follows.
In \S\ 2, we briefly describe the properties of the central caustics induced by a planet and a wide binary companion.
In \S\ 3, we study the central perturbation patterns of caustic-crossing high-magnification events caused by the planetary and wide binary companions.
We also study the effect of the finite source on the planetary central perturbations of the caustic-crossing high-magnification events.
In \S\ 4, we discuss the binary systems that can mimic the planetary central perturbations of the caustic-crossing high-magnification events.
In \S\ 5, we apply the central perturbation patterns of caustic-crossing planetary and binary events to observed events.
We summarize the results and conclude in \S\ 6.

\section{CENTRAL CAUSTIC}

The patterns of the central perturbations in high-magnification events depend on the shape of the central caustics induced by a planet and a very close or a very wide binary companion.
The central caustic induced by the planet has an asymmetric arrow shape for which the horizontal width defined by the separation between the on-axis cusps is longer than the vertical width defined by the separation between off-axis cusps.
The tip of the arrow-shaped caustic points toward the planet.
The horizontal and vertical widths are respectively expressed by
\begin{equation}
\delxip \simeq {4\qp \over {(\spp - 1/\spp)^2}}, \quad
\end{equation}
\begin{displaymath}
\deletap \simeq {\delxip}{{(\spp - \spp^{-1})^{2}|\sin^{3}\phi|\over{( \spp + \spp^{-1} -2\cos{\phi})^{2}}}}\ ,
\end{displaymath}
where
\begin{displaymath}
\qp = {m_p\over{m_1} },
\end{displaymath}
\begin{displaymath}
\cos{\phi} = {{3\over 4}(\spp + \spp^{-1})\left\lbrace 1 - \sqrt{1 - {32\over 9}{1\over{(\spp + \spp^{-1})^{2}}}} \right\rbrace},
\end{displaymath}
where $\spp$ is the projected primary-planet separation normalized by the Einstein radius corresponding to the total mass of the lens system, $\thetae$, and $m_1$ and $m_p$ are the masses of the primary star and planetary companion, respectively \citep{chung05}.
Thus, the vertical/horizontal width ratio, $\Rp$, is represented by
\begin{equation}
\Rp \simeq  {(\spp - \spp^{-1})|\sin^{3}{\phi}|\over{(\spp + \spp^{-1} - 2\cos{\phi})^{2}}} \ .
\end{equation}
According to Equation (2), the shape of the central caustic becomes more asymmetric as $\spp \rightarrow 1$ \citep{chung11}.

On the other hand, the central caustic induced by a wide binary companion has a symmetric asteroid shape.
The horizontal and vertical widths of the central caustic are respectively expressed as
\begin{equation}
\delxib \simeq {4\gamma \over \sqrt{1 - \gamma}}, \quad \deletab \simeq {4\gamma \over \sqrt{1 + \gamma}}\ ;
\end{equation}
\begin{displaymath}
\gamma = {\qb \over{\sbhat^2}},\quad \qb  = {m_{2}\over{m_1}},
\end{displaymath}
where $\gamma$ is the shear induced by the binary companion, $\sbb$ is the projected primary-companion separation normalized by $\thetae$ and $m_2$ is the mass of the companion star, respectively (Chang \& Refsdal 1979; Dominik 1999; Lee et al. 2008).
Here, the $\rm ``hat"$ notation represents the length scale in units of the Einstein radius of the primary, $\thetaeone = \thetae [m_{1}/(m_{1}+m_{2})]^{1/2}$.
Thus, the vertical/horizontal width ratio, $\Rb$, is represented by
\begin{equation}
\Rb \simeq {\left( {1 - \gamma}\over{1 + \gamma}\right )^{1/2}}.
\end{equation}
Because $\sbb \gg 1$, $\gamma \ll 1$ and thus the width ratio becomes $\Rb \sim 1$.

\section{CENTRAL PERTURBATION PATTERNS}

\subsection{\it Caustic-Crossing Events}

To investigate the central perturbation patterns of caustic-crossing high-magnification events caused by a planet and a wide binary companion, we construct magnification excess maps of the planetary and wide binary lens systems.
The magnification excess is defined by
\begin{equation}
\epsilon = {A - A_{0} \over {A_0}}\ ,
\end{equation}
where $A$ and $A_0$ are the lensing magnifications with and without a companion, respectively.

The top panel of Figure 1 shows the magnification excess maps of a planetary system and a wide binary system.
In each map, blue and red color regions represent the areas where the excess is negative and positive, respectively.
The color changes into darker scales when the excess is $|\epsilon| = 1\%,\ 4\%,\ 8\%,\ 16\%,\ 32\%,\ 64\%$, and $96\%$, respectively.
Considering that the masses and distances of the lens of the observed high-magnification planetary-lensing events are mostly distributed in the range of $0.3\ M_{\odot} \leqslant M_{\rm L} \leqslant 0.67\ M_{\odot}$ and $1.0\ \rm{kpc} \leqslant D_{\rm L} \leqslant 6.1\ \rm {kpc}$, respectively, we assume that $M_{\rm L} = 0.5\ M_{\odot}$, $D_{\rm L} = 3\ \rm{kpc}$, $D_{\rm S} = 8\ \rm{kpc}$, and the source is a main-sequence star with a radius of $R_\star = 1.0\ R_\odot$.
Then, the corresponding angular Einstein radius is $\thetae = 0.92\ \rm{mas}$, and the source radius normalized to the Einstein radius is $\rho_{\star} = \theta_\star/\thetae = (R_\star/D_{\rm S})/\thetae = 6.0\times10^{-4}$.

As shown in Figure 1, the excess regions inside the central caustics induced by a planet and a binary companion are differently formed due to the different shape of the two caustics.
Specifically, a notable difference between the two excess regions is the shape of the negative excess region that is formed around the center of the caustic and occupies a significant region inside the caustic.
The shape of the negative excess region for the planetary system is an elongated oval toward the planet and each oval-shaped excess contour has a different center, whereas the shape for the wide binary system is a circle and each circle-shaped excess contour has the same center.
The middle and bottom panels of Figure 1 show the light curves and residuals from the single-lensing event resulting from the source trajectories presented in the planetary and binary maps, in which all the source trajectories pass the negative excess region.

From the analysis of the maps and panels, we find that because of the different negative excess patterns inside the central caustics induced by a planet and a binary companion, the light curves of the caustic-crossing planetary-lensing events exhibit a feature that is discriminated from those of the caustic-crossing binary-lensing events, and the feature can be used to immediately distinguish between the planetary and binary companions.
The planetary-lensing feature appears in the interpeak region between the two peaks of the caustic-crossings.
The structure of the interpeak region for the planetary-lensing events is smooth and convex or boxy, whereas it is smooth and concave for the binary-lensing events.
Since the source passes the negative excess region, the planetary and binary interpeak regions are all negative in the residuals. 
In the planetary-lensing events, the height of the convex feature in the interpeak region decreases as the source trajectory goes away from the center of the caustic.
The convex feature thus disappears in the interpeak region when the source passes a position away from the caustic center (see the planetary-lensing event for trajectory V).
For the planetary-lensing events where the source passes close to the star-planet axis, the lensing light curves do not also exhibit the convex feature in the interpeak region, and they exhibit two very asymmetric peaks of the caustic-crossings, as shown in the planetary-lensing event for trajectory III.
This is because the distances between the first and second caustic-crossing points from the caustic center are considerably different due to a very asymmetric shape of the central caustic induced by a planet.
Thus, the slope between the two peaks of the caustic-crossings for the planetary-lensing events strongly depends on the source trajectory.
On the other hand, the slope between the two caustic-crossing peaks for the binary-lensing events weakly depends on the source trajectory due to a symmetric shape of the central caustic induced by a wide binary companion.
As a result, the slope for the binary-lensing events is almost similar without regard to the source trajectory, as shown in the bottom panel of Figure 1.

\subsection{\it Finite-Source Effect}

The central perturbation patterns in the light curves of the caustic-crossing planetary- and binary-lensing events in Figure 1 are different from each other, but the residual patterns of the central perturbations are almost the same.
If these events are severely affected by the finite-source effect, it becomes more difficult to distinguish the two kinds of caustic-crossing events because the central perturbations of the events would be mostly buried in the light curves.
Therefore, the planetary-lensing feature in the caustic-crossing high-magnification events is significantly affected by the finite-source effect. 
We here investigate the finite-source effect on the planetary-lensing feature in the caustic-crossing high-magnification events.
Since the effect of limb darkening of the source surface is not negligible for the cases when the source diameter is similar to the caustic size, one considers the limb darkening effect for all the cases.
For the limb darkening effect, we adopt a brightness profile for the source star of the form
\begin{equation}
{I(\theta)\over{I_0}} = 1 - \Gamma \left(1-{3\over{2}}{\rm cos}\theta \right ) - \Lambda \left(1-{5\over{4}}{\rm cos}^{1/2}\theta \right ) ,
\end{equation}
where $\Gamma$ and $\Lambda$ are the linear and square-root coefficients and $\theta$ is the angle between the normal to the surface of the source star and the line of sight \citep{an02}.
We also adopt the coefficients of $\Gamma = 0.52$ and $\Lambda = 0.08$ for a main-sequence star.

Figure 2 shows the magnification excess maps of planetary systems with various planet/primary mass ratios and ratios of the caustic to the diameter of the source.
Here, one compares the source diameter to the vertical width of the central caustic induced by a planet, $\deletap$.
In the figure, the projected primary-planet separation is marked out in each map.
As the separation increases, the vertical/horizontal width ratio of the central caustic becomes close to 1 and thus the shape of the caustic gradually becomes symmetric.  
Figure 3 shows the light curves and residuals from the single-lensing event resulting from the source trajectories presented in Figure 2.
Double-peaked central perturbations in the light curves become washed out as the caustic/source size ratio decreases.
The convex structure of the interpeak region in the double-peaked perturbations also become washed out as the caustic/source size ratio decreases.
As a result, we find that the convex-shaped interpeak structure in caustic-crossing planetary-lensing events appears in a certain range, and the range varies depending on the planet/primary mass ratio, i.e., 
\begin{equation}
{\deletap \over {2\rho_\star}} \gtrsim 3.0\sqrt{\qp\over{10^{-3}}}.
\end{equation}
The interpeak region of the caustic-crossing planetary-lensing events where $\deletap/{2\rho_\star} < 3.0\sqrt{\qp/10^{-3}} $ appears as a boxy, a concave, or almost a buried structure. 
For caustic-crossing planetary-lensing events with the concave or the almost buried interpeak structure, it is very difficult to discriminate them from caustic-crossing binary-lensing events.

Since current microlensing follow-up observations intensively monitor high-magnification events and the photometric error reaches $~ 1\%$ at the peaks of the events, the planetary-lensing feature of caustic-crossing high-magnification events can be readily detected.
Therefore, the good coverage of the interpeak region in the caustic-crossing high-magnification events is very important for the discrimination between the planetary and binary companions.

\section{DISCUSSION}

The above planetary-lensing feature arises due to the asymmetric shape of the central caustic induced by a planet.
Since the shape of the central caustic induced by a widely separated binary companion becomes asymmetric as the separation and mass ratio of the binary decrease, it is possible to produce the planetary-lensing feature for the binary-lensing events.
Figure 4 shows the magnification excess map of the binary system with the very low mass ratio of $\qb = 0.01$ and separation of $\sbb = 2.1$.
The central caustic induced by the binary companion has an asymmetric arrow shape like that induced by a planet, as shown in the figure.
However, even if the shape of the caustic is asymmetric, the binary-lensing events passing the negative excess inside the caustic do not produce the convex feature in the interpeak region of the light curves.
This means that the caustic is not asymmetric enough to produce the convex-shaped interpeak structure.
The convex feature in the interpeak region appears in the range of $\Rp < 0.6$ for the planetary-lensing events with the weak finite-source effect (e.g., $\deletap/{2\rho_\star} \gtrsim 5.0$), as shown in Figure 3.
This implies that for the wide binary systems with the severe planet-binary degeneracy, $\Rb \sim 1.0$ and thus the events caused by the wide binary systems do not produce the convex-shaped interpeak structure.

However, we note that the binary systems with the very low mass ratio and separation of $\sbb \rightarrow 1$ can produce the convex-shaped interpeak structure.
In this case, since the size and shape of the caustics of the binary systems are similar to those of the caustics of the planetary systems with massive planets, it would be difficult to distinguish between the binary- and planetary-lensing events with the convex-shaped interpeak structure. 
In addition, the binary systems with $\sbb \sim 1$ can also produce the convex-shaped interpeak structure because of the asymmetric shape and big size of the caustic induced by a binary companion.
However, the events caused by the binary systems due to the big caustic size generally produce much longer caustic-crossing timescales than those of the planetary-lensing events and the overall shape of the binary-lensing light curves including the magnification is different from that of the planetary-lensing light curves.
We thus can easily discriminate two kinds of events through the empirical insight.

\section{APPLICATION TO OBSERVED EVENTS}

We apply the planetary- and binary-lensing features found by this work to observed caustic-crossing high-magnification events with well-covered two peaks.
We note that although an observed event is a caustic-crossing event with two peaks, if the interpeak structure between the two peaks shows a different pattern from those of the planetary- and binary-lensing features and thus it is difficult to apply the two features to the event, we do not consider the event.
For example, MOA-2010-BLG-349 \citep{shin12} is the caustic-crossing event with two peaks where the source with the comparable size to the caustic crosses a tip  of the caustic, but the interpeak structure between the two peaks is different from those of the two features and thus the event is excluded in this application.
Application to four groups with a different double-peaked shape is summarized as below.
\begin{enumerate}
\item[1.]
MOA 2002-BLG-33 \citep{abe03} is the caustic-crossing high-magnification event with well-covered two peaks. 
The interpeak structure between the two peaks is smooth and concave like the binary-lensing feature, but the central perturbations are seemed to be affected by a moderately strong finite-source effect where the source diameter is comparable with the central caustic.
In this case, it is very difficult to distinguish whether the event is caused by a planet or a binary companion.
This is because as mentioned in Section 3.2, the planetary-lensing events affected by moderately strong finite-source effects can produce the smooth and concave interpeak structure, such as the binary-lensing events.
Therefore, this event can be caused by both a binary companion and a planet. 
As the result of the detailed modeling carried out by \citet{abe03}, the event has been identified as the binary-lensing event caused by a binary companion.
\item[2.]
MOA-2008-BLG-159 and MOA-2009-BLG-408 \citep{shin12} are the events with the same central perturbation pattern.
The interpeak structures of these events are smooth and concave and the finite-source effect seems not to be strong.
Thus, we can immediately diagnose that the events are caused by a binary companion.
However, the smooth and concave interpeak structure can be produced by the planetary-lensing events with moderately strong finite-source effects, so it is possible to be considered that they are caused by a planet.
The interpeak structures of the events are actually similar to those of the planetary-lensing events presented in Figure 3.
\citet{shin12} reported that the two events are caused by a binary companion, not a planet through the detailed modeling.
\item[3.]
OGLE 2006-BLG-119, OGLE 2006-BLG-277, OGLE 2006-BLG-284, OGLE 2008-BLG-118, and OGLE 2008-BLG-243 \citep{jaroszynski10} are the events with a smooth and concave interpeak structure. In these events, the perturbations look like very weakly affected by finite-source effects and thus one does not need to consider the finite-source effect.
The caustic-crossing timescales of the five events are over 14 days, and it is much longer than those of observed central caustic-crossing events caused by a planet that have timescales of less than 6 hours (Gould et al. 2006; Dong et al. 2008; Janczak et al. 2010).
The timescale of above 14 days is also much longer than those of the binary events MOA 2002-BLG-33, MOA-2008-BLG-159, and MOA-2009-BLG-408 that have timescales of $\sim 16$ hours, $\sim 2$ days, and $\sim 10$ hours, respectively.
This means that the five events are normal binary-lensing events without the planet-binary degeneracy.
Therefore, we diagnose that these events are caused by a binary companion, and it corresponds to the result of the detailed modeling carried out by \citet{jaroszynski10}.
\item[4.]
OGLE 2006-BLG-076, OGLE 2006-BLG-215, and OGLE 2006-BLG-450 \citep{jaroszynski10} are the events with a convex-shaped interpeak structure.
Although the convex-shaped interpeak structure belongs to the planetary-lensing feature, the caustic-crossing timescales of these events are more than 10 days, so it could be considered that the events are caused by a binary companion.
In addition, the binary systems with $\sbb \sim 1$ can also produce the convex-shaped interpeak structure because of the asymmetric shape and big size of the caustic induced by the binary companion, as mentioned in Section 4.
On the basis of these facts, we diagnose that the three events are caused by the binary systems with $\sbb \sim 1$, and it also corresponds to the result of the detailed modeling carried out by \citet{jaroszynski10}.
\end{enumerate}
In addition to the above events, we have searched other caustic-crossing events that have not yet been reported, but among them there have been no events with a smooth and convex interpeak structure like the planetary-lensing feature.

\section{CONCLUSION}

We have investigated caustic-crossing high-magnification events caused by a planet and a wide binary companion.
From this study, we found that because of the different negative excess patterns inside the central caustics induced by the planet and the binary companion, the light curves of the caustic-crossing planetary-lensing events exhibit a feature that is discriminated from those of the caustic-crossing binary-lensing events, and the feature can be used to immediately distinguish between the planetary and binary companions.
The planetary-lensing feature appears in the interpeak region between the two peaks of the caustic-crossings.
The structure of the interpeak region is smooth and convex or boxy for the planetary-lensing events, while it is smooth and concave for the binary-lensing events.
We have also investigated the finite-source effect on the planetary-lensing feature in caustic-crossing high-magnification events.
As a result, we find that the convex-shaped interpeak structure appears in a certain range of $\deletap/{2\rho_\star} \gtrsim 3.0\sqrt{\qp/10^{-3}} $.

\begin{figure}[t]
\epsscale{1.0}
\plotone{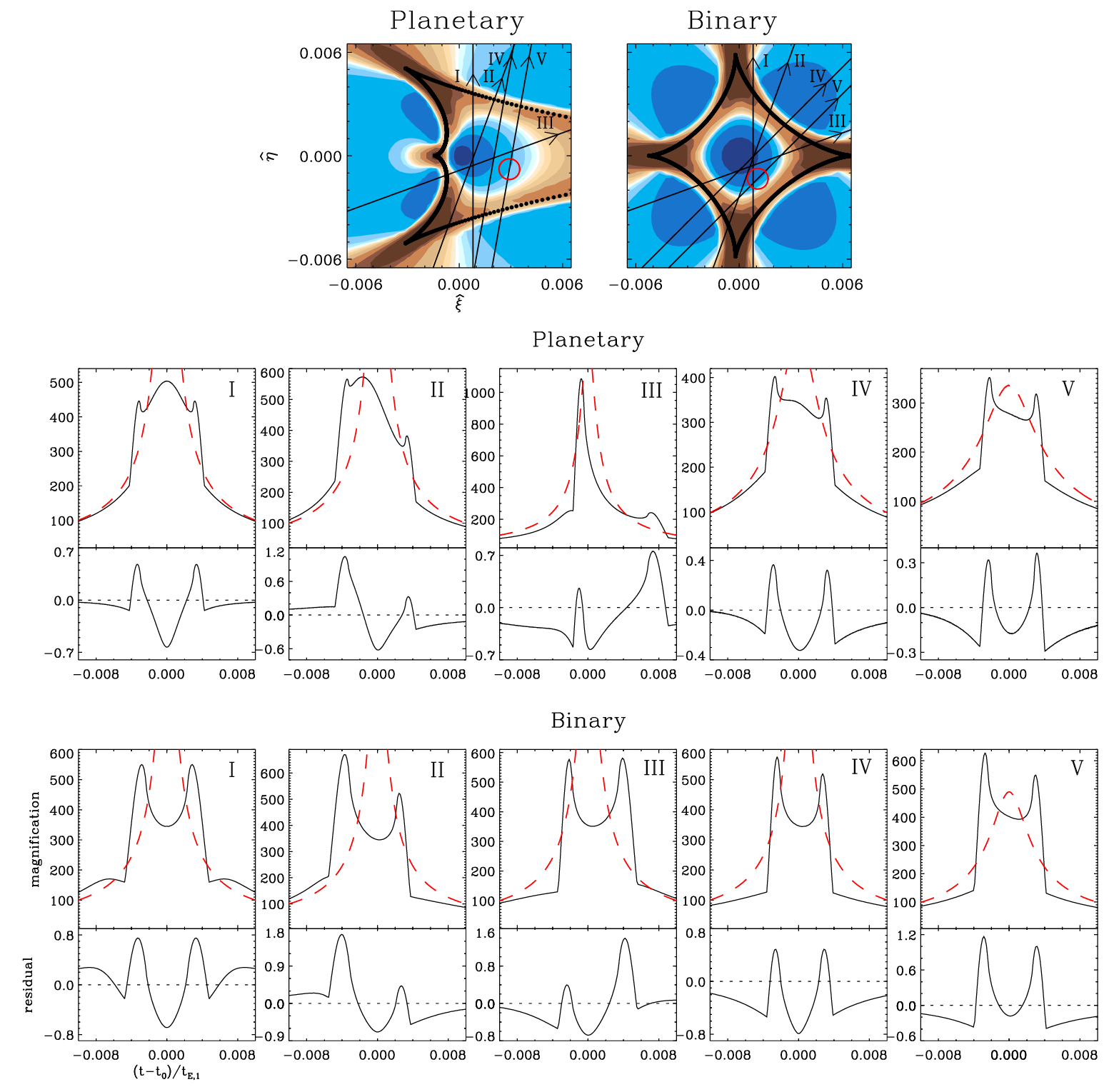}
\caption{\label{fig:one}
Top : Magnification excess maps of a planetary system and a wide binary system.
The lensing parameters of the planetary system are ($\qp,\spp$) = ($3\times10^{-3}, 1.4$), while those of the binary system are ($\qb,\sbb$) = ($0.5, 10.7$).
In the map, the planet and binary companions are located on the right.
The coordinates ($\hat{\xi},\hat{\eta}$) represent the axes that are parallel with and normal to the planetary and binary axes and are centered at the effective position of the primary star.
Here, the notation with the hat represents the length scale normalized by the Einstein radius of the primary, $\thetaeone$.
In each map, the solid circle represents the diameter of the source normalized by $\thetaeone$.
The straight lines with arrows represent the source trajectories.
Middle and Bottom : Light curves and residuals from the single-lensing event resulting from the source trajectories presented in the planetary and binary maps, respectively.
In the middle and bottom panels, the solid curve represents the light curves of the planetary- or the binary-lensing events and the dashed curve represents the light curves of the single-lensing events.
In the residuals, the horizontal line indicates the magnification excess of $|\epsilon| = 0.0$.
}
\end{figure}

\begin{figure}
\epsscale{1.0}
\plotone{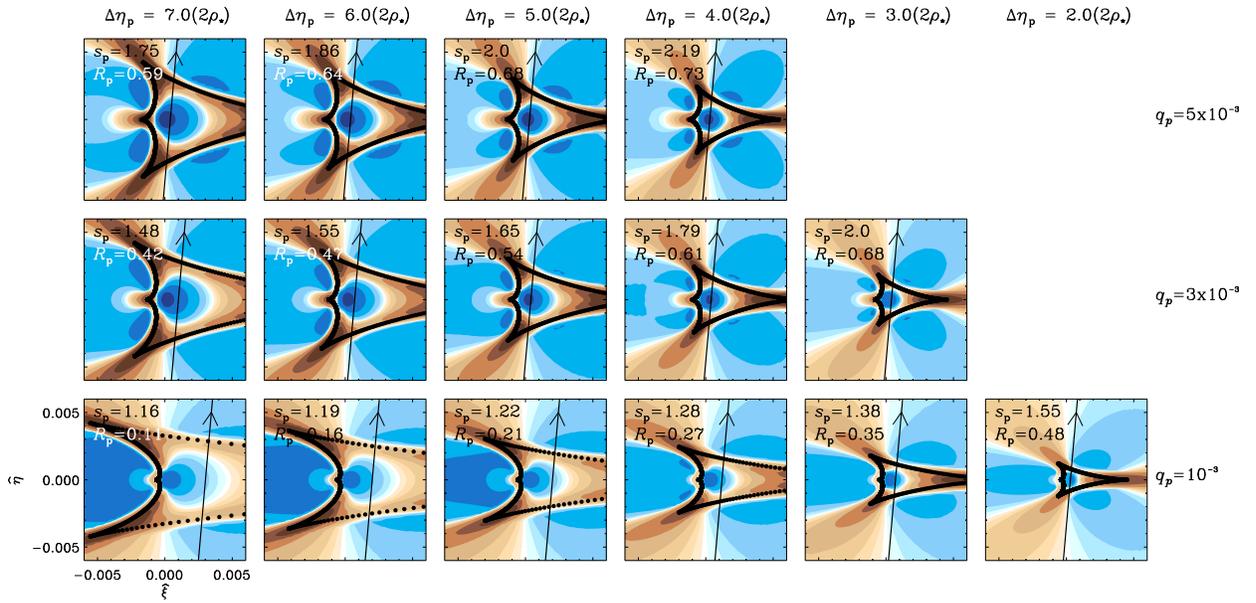}
\caption{\label{fig:two}
Magnification excess maps of planetary systems for various planet/primary mass ratios and ratios of the caustic induced by a planet to the diameter of the source.
In each map, $\Rp$ is the vertical/horizontal width ratio of the central caustic induced by the planet.
}
\end{figure}

\begin{figure}[t]
\epsscale{1.0}
\plotone{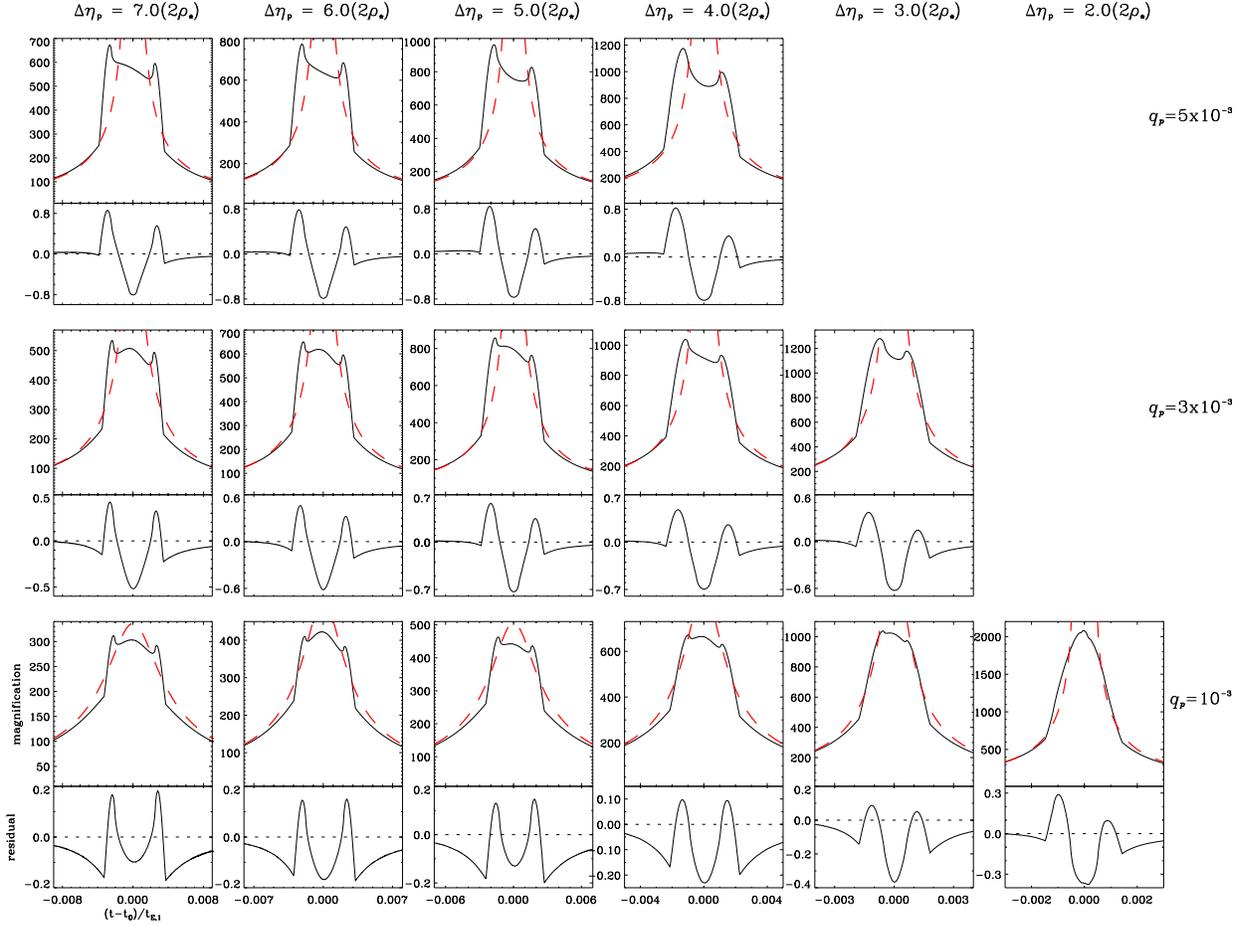}
\caption{\label{fig:three}
Light curves and residuals for the source trajectories presented in Figure 2.
In the upper panel, solid and dashed curves represent the light curves of the planetary- and single-lensing events, respectively.
}
\end{figure}

\begin{figure}[t]
\epsscale{1.0}
\plotone{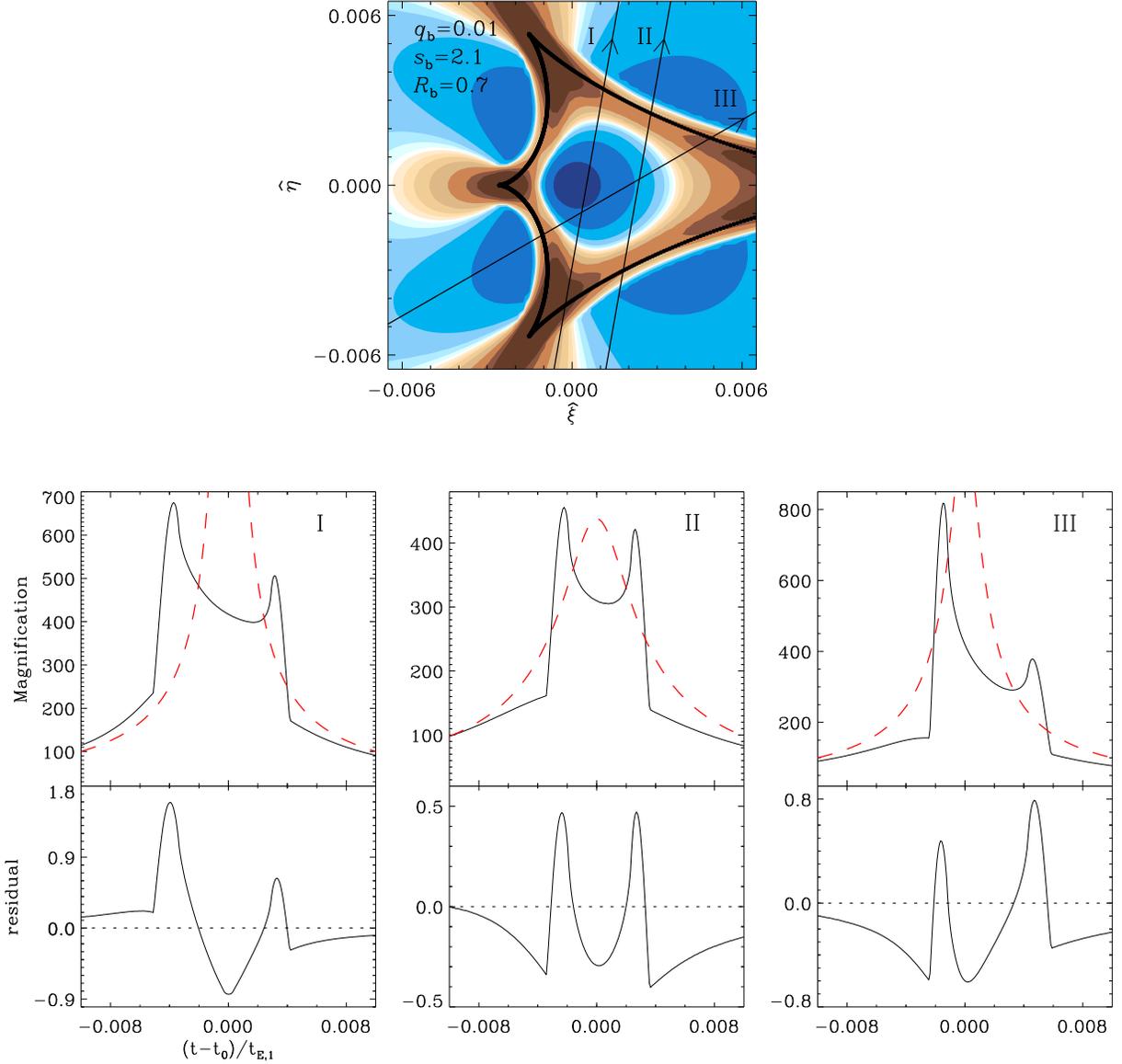}
\caption{\label{fig:four}
Magnification excess map of the binary system with a very low mass ratio together with the light curves and residuals for the source trajectories presented in the map.
In the map, $\Rb$ is the vertical/horizontal width ratio of the central caustic induced by the binary companion.
}
\end{figure}

\acknowledgments
This work was supported by the KASI (Korea Astronomy and Space Science Institute) grant 2012-1-410-02.


\begin{thebibliography}{99}
\frenchspacing

\bibitem[Abe et al.(2003)]{abe03}
Abe, F., Bennett, D.\ P., Bond, I.\ A., et al.\ 2003, A\&A, 411, L493

\bibitem[Albrow et al.(2001)]{albrow01}
Albrow, M.\ D., An, J., Beaulieu, J.-P., et al.\ 2001, \apj, 556, L113

\bibitem[Albrow et al.(2002)]{albrow02}
Albrow, M.\ D., An, J., Beaulieu, J.-P., et al.\ 2002, \apj, 572, 1031

\bibitem[An et al.(2002)]{an02}
An, J.\ H., et al.\ 2002, \apj, 572, 521

\bibitem[Bennett et al.(2008)]{bennett08}
Bennett, D.\ P., Bond, I.\ A., Udalski, A., et al.\ 2008, \apj, 684, 663

\bibitem[Bennett et al.(2010)]{bennett10}
Bennett, D.\ P., Rhie, S.\ H., Nikolaev, S., et al.\ 2010, \apj, 713, 837

\bibitem[Bond et al.(2002)]{bond02}
Bond, I.\ A., Abe, F., Dodd, R.\ J., et al.\ 2002, \mnras, 331, L19

\bibitem[Burgdorf et al.(2007)]{burgdorf07}
Burgdorf, M.\ J., Bramich, D.\ M., Dominik, M., et al.\ 2007, \planss, 55, 582

\bibitem[Chang \& Refsdal(1979)]{chang79}
Chang, K., \&  Refsdal, S.\ 1979, Nature, 282, 561

\bibitem[Chung et al.(2005)]{chung05}
Chung, S.-J., Han, C., Park, B.-G, et al.\ 2005, \apj, 630, 535

\bibitem[Chung \& Lee(2011)]{chung11}
Chung, S.-J., \& Lee, C.-U.\ 2011, \apj, 741, 118

\bibitem[Chung \& Park(2010)]{chung10}
Chung, S.-J., \& Park, B.-G.\ 2010, \apj, 713, 865

\bibitem[Dominik(1999)]{dominik99}
Dominik, M.\ 1999, A\&A, 349, 108

\bibitem[Dong et al.(2009)]{subo09}
Dong, S., Bond, I.\ A., Gould, A., et al.\ 2009, \apj, 698, 1826 

\bibitem[Dong et al.(2006)]{dong06}
Dong, S., DePoy, D.\ L., Gaudi, B.\ S., et al.\ 2006, \apj, 642, 842

\bibitem[Dong et al.(2008)]{dong08}
Dong, S., Gould, A., Udalski, A., et al.\ 2008, \apj, 695, 970

\bibitem[Gaudi et al.(2008)]{gaudi08}
Gaudi, B.\ S., Bennett, D.\ P., Udalski, A., et al.\ 2008, Science, 319, 927

\bibitem[Gaudi \& Gould(1997)]{gaudi97}
Gaudi, B.\ S., \& Gould, A.\, 1997, \apj, 482, 83

\bibitem[Gaudi et al.(1998)]{gaudi98}
Gaudi, B.\ S., Naber, R.\ M., Sackett, P.\ D.\ 1998, \apj, 502, L33

\bibitem[Gould et al.(2006)]{gould06}
Gould, A., Udalski, A., An, D., et al.\ 2006, \apj, 644, L37

\bibitem[Griest \& Safizadeh(1998)]{griest98}
Griest, K., \& Safizadeh, N.\ 1998, \apj, 500, 37

\bibitem[Han \& Gaudi(2008)]{han08}
Han, C.,\& Gaudi, B.\ S.\ 2008, \apj, 689, 53

\bibitem[Janczak et al.(2010)]{janczak10}
Janczak, J., Fukui, A., Dong, S., et al.\ 2010, \apj, 711, 731

\bibitem[Jaroszy\'{n}ski et al.(2010)]{jaroszynski10}
Jaroszy\'{n}ski, M., Skowron, J., Udalski, A., et al.\ 2010, Acta Astron., 60, 197

\bibitem[Lee et al.(2008)]{lee08}
Lee, D.-W., Lee, C.-W., Park, B.-G., et al.\ 2008, \apj, 672, 623

\bibitem[Miyake et al.(2011)]{miyake11}
Miyake, N., Sumi, T., Dong, S., et al.\ 2011, \apj, 728, 120

\bibitem[Shin et al.(2012)]{shin12}
Shin, I.-G., Choi, J.-Y., Park, S.-Y., et al.\ 2012, \apj, 746, 127

\bibitem[Udalski(2003)]{udalski03}
Udalski, A.\ 2003, Acta Astron., 53, 291

\bibitem[Udalski et al.(2005)]{udalski05}
Udalski, A., Jaroszy\'{n}ski, M., Paczy\'{n}ski, B., et al.\ 2005, \apj, 628, L109



\end{thebibliography}
\end{document}